# Hardware Support for QoS-based Function Allocation in Reconfigurable Systems

Michael Ullmann, Wansheng Jin, Jürgen Becker
*Universität Karlsruhe (TH), Germany*
{ullmann, jin, becker}@itiv.uni-karlsruhe.de

## Abstract

*This contribution presents a new approach for allocating suitable function-implementation variants depending on given quality-of-service function-requirements for run-time reconfigurable multi-device systems. Our approach adapts methodologies from the domain of knowledge-based systems which can be used for doing run-time hardware/software resource usage optimizations.*

## 1. Introduction

During the last four years FPGAs have become the favorite prototyping devices in many application areas of computer sciences and electrical engineering. Since modern FPGAs have a higher integration density offering features like partial run-time reconfiguration (e.g. on Xilinx Virtex II FPGAs) they have become attractive for a variety of embedded applications and scientific approaches exploiting theses features. Partial run-time reconfiguration enables for a new class of embedded applications utilizing FPGA resources at run-time as flexible and adaptive hardware-accelerated coprocessors [4]. Furthermore there already exist first academic approaches implementing a complete reconfigurable system-on-chip supporting run-time reconfiguration of dedicated functions and their management at run-time [7][8]. Other researchers already have implemented low budget embedded operating systems running on soft-core or hard-wired on-chip processors on FPGA (e.g. uClinux on a Xilinx MicroBlaze [9]). Combining these approaches using one or several low-cost reconfigurable devices plus dedicated hardware like ASICs or DSPs will create flexible and highly adaptive multi-purpose systems which can be applied in a variety of application domains (e.g. automotive infotainment, multimedia, control-oriented applications etc.). The development and proof of such a versatile system concept is a main research topic of our research group. Our previous work consisted in the development and implementation of a first run-time reconfigurable system-on-chip, supporting flexible on-demand hardware-task switching and a sophisticated run-time reconfiguration and task management mechanisms on Xilinx Virtex II FPGAs [7][10]. Although the tested application domain in our previous work targets at automotive control applications with soft time and security constraints we intend to extend our approach for other fields of application as already mentioned above. Common embedded systems usually have a set of sub-function realizations targeting only one type of hardware for their execution (e.g. as slow software or hardware accelerated functionality only). Additionally the location for execution is normally pre-defined at design time. We address this weak points, since we believe that run-time reconfigurable systems in combination with dedicated hardware resources will have benefits compared to ordinary embedded application approaches which cannot flexibly adapt to changing needs of users. Furthermore by applying intelligent management mechanisms we conceive to gain increases of system-performance and energy/power-efficiency.

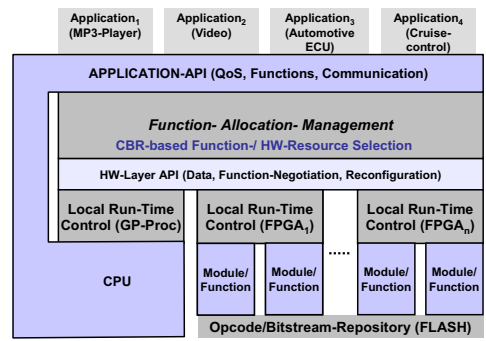

**Figure 1. Reconfigurable HW/SW System**

As shown in figure 1 our previous approach can be re-used as multiple entity at the lower reconfigurable hardware layer of our conceived system approach. Additionally dedicated hardware can be added in parallel. The system is logically divided into different layers. At the application level different applications are executed depending on the location and the mode of operation (as parallel hardware or sequential software tasks). Most applications are conceived to have major parts in software and some dedicated parts accelerated in reconfigurable hardware or DSP. The application level is separated from the lower system levels by an Application-API which offers services for communication, sub-function calls and quality of service (QoS) negotiation. A further system level below is responsible for the proper allocation of functions. Depending on the QoS demands, given by the application's function call an appropriate implementation of the desired function must be found from a run-time function repository. So this layer needs informations about the available functions, their different implementations and QoS features. Additionally it will need informations about the current system load and power consumption status, which are procured by the HW-Layer API one level below. The HW-Layer API is the interface for all hardware relevant aspects like resource consumption, low-level communication and reconfiguration of system parts. It connects the high level components with the local system controllers, which may be located on different devices (e.g. standard CPU, FPGA (soft-core CPU) or DSP). These controllers are responsible for the control of local run-time reconfiguration and other sub-tasks like local task/ resource management and communication issues.

This paper will focus on some aspects of a QoS-aware function allocation so that details on the other system levels as they were not published yet will be presented in future papers. Our contribution is structured as follows: In section 2 we give a motivation for QoS-aware function allocation and how it can be solved by means of a simplified case-based-reasoning approach. Section 3 describes a short application example for our case-based-reasoning approach whereas section 4 describes the hardware/ software implementation and synthesis results of the algorithm. The paper closes with a summary and gives an outlook on our future work.

## 2. QoS-aware Function Allocation

In some cases, especially in multimedia applications it is not sufficient to do a simple function call if a dedicated function is needed which has to comply with additional constraints like data/frame-rates, processing modes, response deadlines etc. We conceive that the system offers for one requested function type different implementations which can be run as software or as reconfigurable accelerated hardware having different features. So there emerges the problem to identify a set of most appropriate implementation variants which match best to the given constraints from request. The found set of implementation variants can be used for checking the current system load and resource consumption state concerning the feasibility of a best matching implementation out of it which can be inserted on FPGA or as software-task on a processor. It is possible that the best matching implementation is not currently feasible without preempting other active (hardware) tasks so an alternative implementation can be offered to the calling application which has to decide on it. The details on this QoS negotiation mechanism as they are not presented here will be in the scope of future papers. At this point it is primarily here of interest how a best matching implementation by given constraints can be found. Before we introduce our approach we want to give an overview on case-based reasoning and how parts of this approach can be applied for solving that problem.

### 2.1. Case-based Reasoning Background

Case-based reasoning (CBR) is an approach for developing knowledge-based systems that are able to retrieve and reuse solutions that have worked for similar situations in the past. CBR traces its roots to the work of Roger Schank, Janet Kolodner and Michael Lebowitz in the early 1980s [2][5][6]. Since the beginning of the 1990s CBR approaches are extensively applied in help-desk applications and diagnostic expert systems for customer support. In CBR systems expertise is embodied in a library of past cases, rather than being encoded in classical rules. Each case typically contains a description of the problem, plus a solution and/or the outcome.

The knowledge and reasoning process used by an expert to solve the problem is not recorded, but is implicit in the solution. To solve a current problem: the problem is matched against the cases in the case base, and *similar cases* are *retrieved*. The retrieved cases are used to suggest a solution which is reused and tested for success. If necessary, the solution is then revised. Finally the current problem and the final solution are retained as part of a new case. The complete CBR-cycle is shown in figure 2.

### 2.2. Case-base Representation and Similarity

Problem cases may have different representations. These can be object-oriented, trees & graphs or sets of simple pairs of attributes and their values. We have found the latter representation appears to be best suited for our purposes, since attributes of some type may describe comparable features of different implementations. The attributes' values depend on their type and given value range and can be of integer/real type, even discrete ordered sets of symbols are possible if they can be mapped onto integers. Typical types can be data-rates, discrete processing modes (float/integer), power consumption, code/bitstream-sizes, response times, frame sizes, max. bit-error-rates etc. A *local similarity* measure is needed for comparing attributes of same type between different implementations. Such a measure is often based on a transformation function which calculates from the Euclidian or Manhattan distance of two given attributes a similarity value in the range [0 ... 1], where 1 means that both attribute values are identical and 0 means that both values have a maximum distance (no similarity).

$$s_i(x_A, x_B) = 1 - \frac{d(x_A, x_B)}{1 + max(d(x_i, x_j))}; \; d(x_A, x_B) \leq 0. \quad (1)$$

An example function is given in equ. (1) where $x_A$, $x_B$ are attribute values of same type from a request A and an implementation case B. The function $d(x_i, x_j)$ calculates the distance / absolute difference between both values where $max(d(x_i, x_j))$ represents the maximum possible distance which can be determined at design time from all attributes of same type given by the implementation library. Since the request and implementation descriptions may contain several attributes it is not sufficient to calculate for every attribute pair $(A_{Req\_i}, A_{Impl\_i})$ a local similarity $s_i$. All local similarities $s_i$ have to be combined into a *global similarity* which enables for comparing all implementation variants of same basic function type with the attribute description set of a given function request. Such a needed function $S_{global}$ is denoted as *amalgamation function* which transforms an input vector located inside an n-dimensional cube $[0 ... 1]^n$ back into a scalar range of [0 ... 1]. A convenient function is the weighted sum of all local similarities as shown in equ. (2). It is monotonous in every argument and $S_{global}(0, ..., 0)=0$; $S_{global}(1, ..., 1)=1$.

$$S_{global}(s_1, ..., s_n) = \sum_{i=1}^{n} w_i \cdot s_i; \quad \sum_{i=1}^{n} w_i = 1. \quad (2)$$

At this point it should be noted that other approaches for similarity calculations are possible as well. A well known method comes from statistical decision theory and determines the Mahalanobis distance by calculating the co-variance matrix of the whole set of function attributes. This method is very effective concerning the results but the computational efforts would be too large so we decided to apply Manhattan distance metrics.

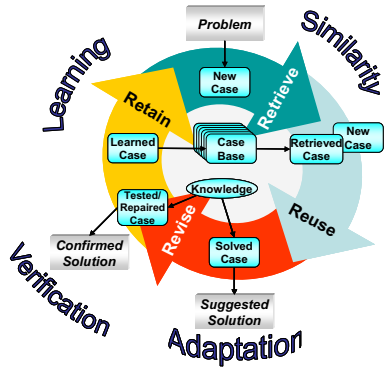

**Figure 2. Case-based reasoning- cycle [1]**

## 3. Application Example for Retrieval

Following short application example will give an impression how a retrieval and similarity comparison can be done. As shown in figure 3 an application requires an FIR-equalizer functionality for audio DSP purposes. Each offered type of basic-functionality has a global function-ID which is used for finding the proper type entry inside the function implementation tree. The nodes of this tree are ordered in a hierarchy, where the nodes in the upper level represent all function types whereas their successor nodes at lower levels contain informations about their related implementations like the implementation-ID which can have a unique system-global or a local ID value and a set of attributes, separable again by unique type IDs, which contain details on each implementation's features like processing bitwidth, processing mode (integer/float), output mode and sampling rate. Other attributes like power consumption, response deadlines etc. are conceivable. It should be noted here that such metrics which characterize a functionality on QoS-aspects have to be pre-defined by the designer as a set of attributes whose values are derived from simulations and tests of the function's model. Depending on the application's needs the request's attribute composition may vary.

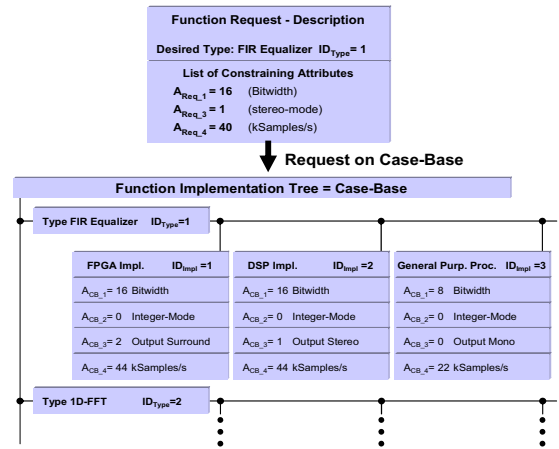

**Figure 3. Function request at case-base**

As first step all function type entries have to be checked for finding the required type ($ID_{Type}$). It should not happen that the desired type is not found since the application's functional requirements should already be known at design time, otherwise the function can not be served. In the given example case the desired function-$ID_{Type} = 1$ is found and three possible implementations for different execution targets (FPGA, DSP, General Purpose Proc.) have to be checked. Figure 3 shows that the request's attribute-set does not have to be completely specified; incomplete subsets are possible as well which is a nice property of case-based retrieval. As next step all corresponding attributes are retrieved from each implementation attribute sub-table and for every implementation $k$ a similarity value $S_{global}(k)$ is calculated by applying equ. (1) and (2) as shown in table 1. If a corresponding implementation attribute is not found, its local similarity $s_i$ can be set to 0 because a missing attribute can be seen as unsatisfiable requirement.

The $d_{max}$ values as used in table 1 were taken from an extra table (not shown here) which was generated at design time containing supplemental data on the attributes' design-global upper/lower value bounds (see also figure 4 (right – $maxrange^{-1}$)).

As the results show from table 1 the DSP-based function-implementation matches best to the given requirements. The FPGA-implementation produces the second-best similarity whereas the standard software implementation has a rather low similarity which would not satisfy the demands if the attributes were inspected manually. It's conceivable to reject all results below a given threshold similarity. In the given example the allocation manager would check now for each acceptable solution its feasibility concerning the system load and would suggest the remaining implementation-variants to the calling application. Since every available function realization has a unique identifier it will be possible to retrieve the function's corresponding configuration data (CPU opcode/ FPGA bitstream) from a global function repository for reconfiguration.

**Table 1. Retrieval – similarity example**

| i | $A_{Req\_i}$ | $A_{CB\_i}$ | $d(A_{Req\_i}, A_{CB\_i})$ | $d_{max}$ | $s_i$ |
|---|---|---|---|---|---|
| 1 | 16 | 16 | 0 | 16-8=8 | 1 |
| 3 | 1 | 2 | 1 | 2-0=2 | 0.66 |
| 4 | 40 | 44 | 4 | 44-8=36 | 0.89 |
| Impl. ID= 1 : FPGA | | | | $w_i$=1/3 | $S_{global}\approx$ 0.85 |
| i | $A_{Req\_i}$ | $A_{CB\_i}$ | $d(A_{Req\_i}, A_{CB\_i})$ | $d_{max}$ | $s_i$ |
| 1 | 16 | 16 | 0 | 16-8=8 | 1 |
| 3 | 1 | 1 | 0 | 2-0=2 | 1 |
| 4 | 40 | 44 | 4 | 44-8=36 | 0.89 |
| Impl. ID= 2 : DSP | | | | $w_i$=1/3 | $S_{global}\approx$ 0.96 ← best |
| i | $A_{Req\_i}$ | $A_{CB\_i}$ | $d(A_{Req\_i}, A_{CB\_i})$ | $d_{max}$ | $s_i$ |
| 1 | 16 | 8 | 8 | 16-8=8 | 0.11 |
| 3 | 1 | 0 | 1 | 2-0=2 | 0.66 |
| 4 | 40 | 22 | 18 | 44-8=36 | 0.51 |
| Impl. ID= 3 : GP-Proc | | | | $w_i$=1/3 | $S_{global}\approx$ 0.43 |

It is still possible that no matching feasible variant was found so that the application has to repeat its request with rather relaxed constraints giving a chance to the third low performance implementation ($ID_{Impl}$=3). Otherwise the application can not call the function. If a function was allocated and instantiated on hardware it is not necessary to repeat the retrieval procedure at repeated function calls. The allocation manager could create a kind of bypass-token containing data one the previous selection which can be reused at repeated function calls so that only an availability check on the function and its allocated resources has to be done.

## 4. Hardware/Software Implementation

The function allocation manager's retrieval functionality can be implemented in software or as hardware mapped algorithm. Although case-based retrieval is a rather control oriented algorithm we have been able to model, simulate and synthesize an accelerated retrieval unit on FPGA.

### 4.1. Data Structures

As first step the needed data structures for request and implementation-tree were defined. We decided to use linear lists which can be connected by reference pointers for creating complex tree structures. Each list contains several entries like IDs, values, pointers and is terminated by a dedicated NULL-entry. These lists can be easily mapped on linear organized RAM-blocks if all list elements use the same word length per entry (e.g. 16 or 32 bits). Figure 4 (left) shows the structure of a list containing the request description including the desired function type, attributes and weighting factors $w_i$ to be used. The internal order of entries is predefined so that an attribute's ID is always followed by its value and weight. Additionally the attribute-blocks have to be pre-sorted by their ID in ascending order. This measure is applied in all other list-structures as well (see figure 5) and aims at improving the retrieval efficiency of the algorithm. Because each attribute has to be searched by its ID in each implementation's attribute-list it is possible to avoid a repeated search from the top of each list.

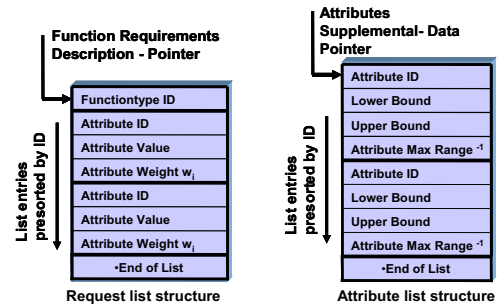

**Figure 4. Request list and attribute list**

Since the next requested attribute has a larger ID value than its predecessor it is possible to continue searching from the current position instead of doing a repeated search from the top of the local list. As a consequence the effort for searching becomes linear.

Another auxiliary list which is used for similarity calculation is shown in figure 4 (right). The entries are grouped again in blocks and they are pre-sorted by attribute IDs for the same reasons as mentioned before. The fourth entry of each attribute block ($maxrange^{-1}$) contains a pre-calculated reciprocal value of $d_{max}+1$.

Since it is a constant we do not need to implement an expensive hardware divider saving resources. By using the reciprocal value we can do a rather fast multiplication with the attributes' absolute difference instead of doing a slow division (see also equ. (1)). Figure 5 shows the implementation-tree structure which is a hierarchical tree of three levels. Similar to figure 3 it contains a top level list including implementation-IDs and reference pointers to the corresponding implementation lists. Each implementation list contains blocks sorted by implementation ID with pointers referencing to lists of attribute/value pairs of each implementation. All partial lists are generated at design time creating one big block of linear concatenated lists.

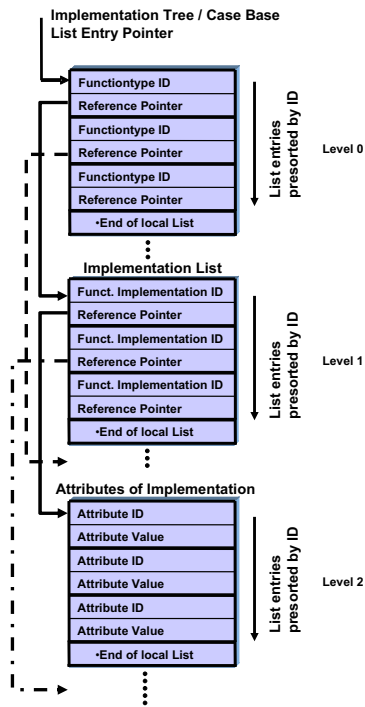

**Figure 5. Implementation-tree structure**

### 4.2. Hardware Implementation

The hardware implementation of the retrieval unit was done by modeling its behavior in Matlab Stateflow at first. We developed some tools in Matlab for creating and exporting all needed data structures (implementation-tree, request list etc.) so that they can be easily used for testing purposes in Stateflow, VHDL and C. After testing and verifying of our Stateflow model we converted the behavioral model into synthesizable VHDL code by using a special conversion tool JVHDLgen [3]. This tool is still in beta state of development but it proved to work fine although we had to do some minor restrictions to our Stateflow model since not all features of Stateflow are currently supported. Additionally some manual code modifications were necessary for synthesizing the model onto Xilinx Virtex II 3000 FPGA using Xilinx ISE 6.2.

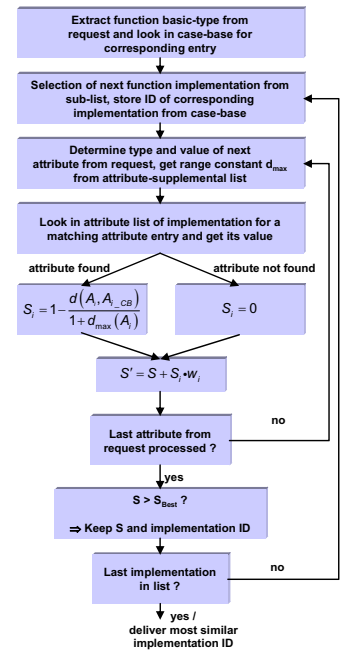

**Figure 6. Most similar retrieval algorithm**

Figure 6 gives an overview on the main parts of the implemented algorithm. The shown version is able to find the most similar implementation. The delivered results will be the ID and similarity value of the best matching implementation. The main components of the retrieval unit's data path are depicted in figure 7. It should be noted that this data path's schematic was derived from the Stateflow model as the generated VHDL code is less suitable for extracting control- and data path information. The processing bitwidth of all attribute values was defined at 16 bit. Our tests showed that this bitwidth is sufficient even for fixed point calculations without seriously losing accuracy. We have been able to show that we get the same retrieval results in high precision floating point Matlab simulation as we get from VDHL simulation using ModelSim.

**Table 2. Synthesis results on XC2V3000**

| Resources: Xilinx Virtex II 3000 | |
|---|---|
| CLB-Slices: | 441 of 14336   ≈ 3 % |
| MULT18X18s: | 2 of 96   ≈ 2 % |
| BRAMS(18Kbit): | 2 of 96   ≈ 2 % |
| Max. Clock: | ≈ 77 MHz |

The hardware design takes 441 CLB slices, two 2x18bit hardware multipliers and can be operated at 75 MHz (see also table 2). A small amount of additional memory of about 4.5 kB is needed for storing the implementation-tree giving space for a full set of 15 function types containing 10 implementations * 10 attributes each, using 16 bit words for each entry (see table 3, reference pointers are included).

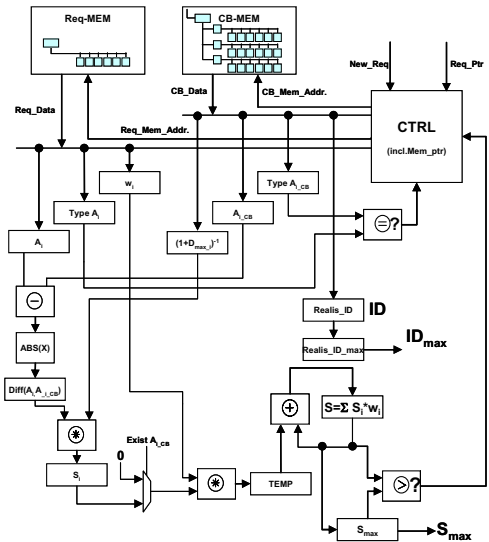

**Figure 7. Data-path - Most similar retrieval**

Apart from the hardware implementation we also mapped the retrieval algorithm into a C program running on a Xilinx MicroBlaze soft-processor at 66 MHz. The software version which takes only 1984 bytes of opcode and 1208 bytes for variables proved to produce identical retrieval and similarity results for a selected set of test cases where we created different implementation-trees and requests. The same test cases were applied to the hardware implementation and we compared the performance results of both implementations. As result we have found that our hardware version is at 66 MHz about 8.5 times faster than the software solution.

**Table 3. Case-base memory consumption**

| Different types of attributes in total: | 10 |
|---|---|
| Implementations per function type: | 6 |
| Attributes per Request: | 10 (worst case) |
| Types of basic functions in total: | 15 |
| Attributes per Implementation: | 10 |
| Memory consumption of case-base: (16 bit-words each entry/pointer) | 4.5 kB |
| Memory consumption of request: | 64 Bytes |

## 5. Conclusions and Outlook

We have proved the feasibility of a hardware accelerated function-retrieval on QoS requirements based on methodologies from case-based reasoning theory. Although we adopted the CBR-retrieval and similarity determination steps for our purposes some might argue that the presented approach does not implement a complete CBR-cycle as shown in figure 2. Actually many practical CBR-implementations restrict to the retrieval step only and re-use the found solution without adaptation and assessment step, since a reasonable adaptation of the found solution is a very complex and time consuming process, which is not necessary in a retrieval of static implementations. Although the implementation-tree is currently a static structure we conceive dynamic update mechanisms of Case-Base-data structures and function repositories at run-time enabling for a self-learning system as new aspects of our future work. Our next step will be an extension for getting n most similar solutions from retrieval which offers the possibility for checking out the feasibility of different matching variants. Additionally we think about optimizations of the used implementation-tree structure and retrieval finite state automaton for getting a better speed-up. Furthermore a rather compacted attribute block representation could be used for loading IDs and values as blocks within one step speeding everything up at least by factor 2.